\documentclass[12pt]{article}
\usepackage{pdproc,epsfig} 

\begin{document}

\renewcommand{\thefootnote}{\alph{footnote}}

%%%%%%%%%%%%%%%%%%%%%%%%%%%%%%%%%%%%%%%%%%%%%%%%%%%%%%%%%%%%
%% General macros and hyphenation directives.
%%%%%%%%%%%%%%%%%%%%%%%%%%%%%%%%%%%%%%%%%%%%%%%%%%%%%%%%%%%%
\newcommand{\etal}      {{\it et al.}}
\hyphenation{Che-ren-kov} 
\hyphenation{Ka-mio-kande} 

%%%%%%%%%%%%%%%%%%%%%%%%%%%%%%%%%%%%%%%%%%%%%%%%%%%%%%%%%%%%
% Define macros to typeset the units.
%%%%%%%%%%%%%%%%%%%%%%%%%%%%%%%%%%%%%%%%%%%%%%%%%%%%%%%%%%%%

\newcommand{\cm}        {\ensuremath{\mbox{cm}}}
\newcommand{\meter}     {\ensuremath{\mbox{m}}}
\newcommand{\mwe}       {\ensuremath{\mbox{mwe}}}
\newcommand{\km}        {\ensuremath{\mbox{km}}}
\newcommand{\kilometer} {\km}
\newcommand{\unit}[1]   {~\mbox{#1}}
\newcommand{\yr}	{\ensuremath{\mbox{yr}}}
\newcommand{\GeV}       {\ensuremath{\mbox{GeV}}} 
\newcommand{\GeVc}      {\ensuremath{\GeV{}\!/\mbox{c}}}
\newcommand{\MeV}       {\ensuremath{\mbox{MeV}}} 
\newcommand{\MeVc}      {\ensuremath{\MeV{}\!/\mbox{c}}}
\newcommand{\MeVcc}      {\ensuremath{\MeV{}\!/\mbox{c}^2}}
\newcommand{\kiloton}   {\ensuremath{\mbox{kt}}}
\newcommand{\kilotonyr} {\ensuremath{\mbox{kt}\!\cdot{}\!\mbox{yr}}}
\newcommand{\megatonyr} {\ensuremath{\mbox{Mt}\!\cdot{}\!\mbox{yr}}}
\newcommand{\degree}    {\ensuremath{^{\circ}}}
\newcommand{\millirad}  {\ensuremath{\mbox{mrad}}}
\newcommand{\inch}      {\ensuremath{^{\prime\prime}}}
\newcommand{\sys}	{~\mbox{(sys.)}}
\newcommand{\stat}	{~\mbox{(stat.)}}
\newcommand{\theo}	{~\mbox{(theo.)}}

%%%%%%%%%%%%%%%%%%%%%%%%%%%%%%%%%%%%%%%%%%%%%%%%%%%%%%%%%%%%
% Define macros to typeset extra math symbols.
%%%%%%%%%%%%%%%%%%%%%%%%%%%%%%%%%%%%%%%%%%%%%%%%%%%%%%%%%%%%

\newcommand{\lessthan}	{\ensuremath{<}}
\newcommand{\greaterthan}{\ensuremath{>}}
\newcommand{\E}[1]{\ensuremath{\times 10^{#1}}}
\newcommand{\lsim}      {\ensuremath{\mathrel{\raisebox{-.6ex}{
        \ensuremath{\stackrel{\textstyle<}{\sim}}}}}}
\newcommand{\gsim}      {\ensuremath{\mathrel{\raisebox{-.6ex}{
        \ensuremath{\stackrel{\textstyle>}{\sim}}}}}}
\newcommand{\about}[1]{\ensuremath{\sim}}

%%%%%%%%%%%%%%%%%%%%%%%%%%%%%%%%%%%%%%%%%%%%%%%%%%%%%%%%%%%%
% Typeset experiment names.
%%%%%%%%%%%%%%%%%%%%%%%%%%%%%%%%%%%%%%%%%%%%%%%%%%%%%%%%%%%%
\newcommand{\sk}        {Super-Kamiokande}
\newcommand{\SK}        {Super-Kamiokande}
\newcommand{\SuperK}    {\sk{}} 
\newcommand{\Frejus}    {Fr\'{e}jus} 
\newcommand{\kt}        {1\unit{\kiloton{}}}

%%%%%%%%%%%%%%%%%%%%%%%%%%%%%%%%%%%%%%%%%%%%%%%%%%%%%%%%%%%%
% Define macros to typeset particles
%%%%%%%%%%%%%%%%%%%%%%%%%%%%%%%%%%%%%%%%%%%%%%%%%%%%%%%%%%%%
\newcommand{\goesto}    {\ensuremath{\rightarrow}}
\newcommand{\nux}       {\ensuremath{{\nu}_{\rm x}}}
\newcommand{\nue}       {\ensuremath{{\nu}_{\rm e}}}
\newcommand{\nuebar}    {\ensuremath{\bar{\nu}_{\rm e}}}
\newcommand{\numu}      {\ensuremath{\nu_{\rm \mu}}}
\newcommand{\numubar}   {\ensuremath{\bar\nu_{\rm \mu}}}
\newcommand{\nutau}     {\ensuremath{{\nu}_{\rm \tau}}}
\newcommand{\nutaubar}  {\ensuremath{\bar{\nu}_{\rm \tau}}}
\newcommand{\nus}       {\ensuremath{{\nu}_{\rm s}}}
\newcommand{\mnu}       {\ensuremath{{\rm m}_\nu}}
\newcommand{\pizero}    {\ensuremath{\pi^{0}}}

\newcommand{\peppo}     {\ensuremath{p \goesto{} e^+ \pi^0}}
\newcommand{\pmppo}     {\ensuremath{p \goesto{} \mu^+ \pi^0}}
\newcommand{\pnukp}     {\ensuremath{p \goesto{} \bar{\nu} K^+}}

\newcommand{\sstt}      {\ensuremath{\sin^2 2\theta}}
\newcommand{\dms}       {\ensuremath{\Delta m^2}}
\newcommand{\eVs}       {\ensuremath{\mbox{eV}^2}}

\title{
 UNO}

\author{ R. JEFFREY WILKES, for the UNO Collaboration}

\address{ Department of Physics, University of Washington\\
 Seattle, WA 98195, USA\\
 {\rm E-mail: wilkes@u.washington.edu}}

\abstract{
UNO (Underground Neutrino Observatory) is intended as  
a  multi-purpose  detector, potentially useful for purposes which may not
have been considered at the time of its construction. UNO accomodates a  comprehensive 
physics program, encompassing nucleon decay and lepton flavor  physics,  including  
the possibility of observing CP
violation,  grand  unification  scale  physics, and detailed observations
of supernova explosions. The conceptual design of UNO is described, along with 
specifics of how UNO would fit into an existing commercial mine site in the 
western USA,
which has been used as a strawman site for planning and proposal development.
 }
   
\normalsize\baselineskip=15pt

\section{Introduction}

Water Cherenkov detectors play the role in neutrino physics that
bubble chambers did in mid-20th century particle physics: large volume
tracking detectors sufficiently unspecialized to adapt to
physics goals that are moving targets. 
Their versatility is emphasized by noting 
that early ring imaging water Cherenkov detectors were originally 
designed to search for nucleon decay, but later became the
weapon of choice for attacking neutrino puzzles.
Capitalizing  on   this
versatility, UNO (Underground Neutrino Observatory) is intended as  
a  multi-purpose  detector, potentially useful for purposes which may not
have been considered at the time of its construction. UNO accomodates a  comprehensive 
physics program, encompassing nucleon decay and lepton flavor  physics,  including  
the possibility of observing CP
violation,  grand  unification  scale  physics, and detailed observations
of supernova explosions.

Recently, neutrino physics has become the most exciting and rapidly-advancing 
area of particle physics
and astrophysics. This is in large part due to results from two large underground 
water Cherenkov detectors, \SK ~and SNO, including the
discovery of neutrino  oscillations  and  neutrino  mass using atmospheric 
neutrinos; resolution of the long-standing solar  neutrino  flux
deficit in terms of oscillations, first
detection  of  accelerator-produced  neutrinos in a long-baseline
experiment, and  establishment  of  the
world's best limits on nucleon decay, excluding
minimal  SU(5)  GUTs and MSSM SU(5). Extending these 
accomplishments by constructing
a next-generation, million ton scale detector like UNO is a widely accepted goal, 
supported by planning 
groups such as (in the USA) the HEPAP sub-panel on Long Range Planning in 2001
\cite{ref:hepaplrp01}, 
the Committee
on the Physics of Universe (CPU) in 2001\cite{ref:cpu01}, the Neutrino Facility Assessment
Committee (NFAC) sponsored by the National Academy of Science in 2002, and
the HEPAP Facilities Committee in 2003\cite{ref:NAS}\cite{ref:Gomez}. 
Parallel to the UNO initiative, 
the possibility of  similar  next-generation
underground water Cherenkov detectors are being discussed in  Japan  (Hyper-
Kamiokande)\cite{ref:HK} and Europe (Memphys at Fr\'{e}jus)\cite{ref:Frejus}. 
These reports reinforce the  broad
endorsement of the physics objectives  UNO  aims  to  address,  and  
demonstrate a  global
commitment to  shared  goals.  

The physics potential of UNO is further enhanced by the recent realization
that CP violation in  neutrino  sector  can  be  measured  using  a
conventional neutrino super-beam and a large water Cherenkov  detector  with
a very long baseline  (2000-3000  km),  utilizing  the  secondary  oscillation
maxima. \cite{ref:Marciano} 
A preliminary study of a super-long baseline experiment using a wide-band
neutrino beam produced by the 
upgraded BNL-AGS accelerator was very  encouraging\cite{ref:BNL}.  
While these results and the details of the plan need  to be
verified by further studies, the BNL proposal suggests
a novel way of measuring neutrino  oscillation  parameters  and  CP
violation using a conventional high flux wide band neutrino beam.

UNO was first proposed at the NNN99 Workshop in September, 1999\cite{ref:NNN}.
An  informal
UNO proto-collaboration was formed in 2000. A comprehensive  study  of  the
physics potential and feasibility of the detector was carried  out  and  the
results were presented in 2001 at the Snowmass Workshop\cite{ref:UNO}.
Since  then, a series of workshops  has been held,  and
planning for a large next-generation water Cherenkov  detector, not tied to any 
particular site,  has
attracted  wide interest\cite{ref:UNOweb}. 
Recognizing the necessity of international 
collaboration for a large project like UNO, we are 
committed to make the collaboration truly international, serving as a
facility for the worldwide physics community.

In 2003, 
a formal collaboration 
was organized to prepare for proposals to 
funding agencies.  At  present,  the UNO  
collaboration  consists  of  96  experimental scientists and engineers,  
representing  37 institutions from 7 countries. 
The collaboration is  supported  by  a  10  member  Theoretical
Advisory Committee, 11 member Advisory Committee composed of experimentalists 
and other interested researchers from Canada, China,  Europe,
Japan, and the United States, numbering about 150 in total. 

While the UNO
collaboration remains independent of any particular site, we have identified
an excellent candidate site in the Henderson Mine, one hour's drive
west of Denver, Colorado, in the 
western USA. Parameters describing 
the Henderson Mine will henceforth be used to provide
concrete results in numerical simulations and planning activities, without
prejudicing future siting options. Features of the Colorado site 
will be described in a later section.

\section{The UNO Detector}

Extension of  the  well-established  water  Cherenkov
technique  to  achieve  an
order of magnitude better sensitivity to nucleon decay and neutrino  physics
presents no serious  technical  challenges.  To  strike  a  balance  between
increased physics reach and practical considerations such as cost, the  nominal
fiducial volume of the UNO detector is taken to be 20 times that of \SK. 
There are 
several  practical
constraints on the size scale of a water Cherenkov detector.  Water  depth  is
limited by the pressure tolerance of existing PMTs ($\sim$8 atm,
for current 20" Hamamatsu PMTs), unless new photosensors are developed. 
The attenuation length of Cherenkov
light in pure water ($\sim$80 m at $\lambda$ = 400 nm in SuperK) sets another limit.
Finally, unsupported cavity widths greater than $\sim$60 m would push the envelope
of underground excavation engineering. 
Studies show that a 
segmented configuration, comprising a linear array of 60 meter cubes,
is the best choice for UNO. Three such segments give a net
water mass of about 650 kT, for example. 

\begin{figure}
\vspace*{13pt}
\leftline{\hfill\vbox{\hrule width 5cm height0.001pt}\hfill}
         \mbox{\epsfig{figure=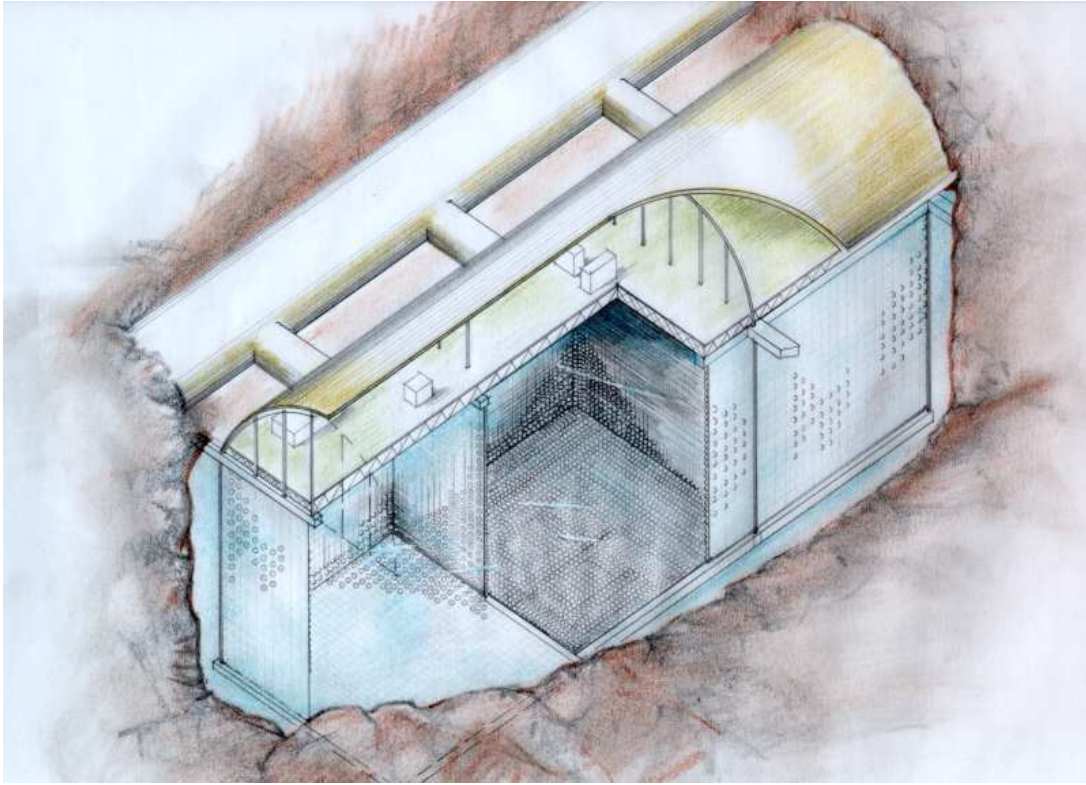,width=0.95\textwidth}}
\vspace*{1.4truein}		%ORIGINAL SIZE=1.6TRUEIN x 100% - 0.2TRUEIN
\leftline{\hfill\vbox{\hrule width 5cm height0.001pt}\hfill}
  \caption{\label{fig:UNODetector} Baseline conceptual design of UNO showing the
    central detector module (40\% photo-cathode coverage) with the outer
    wing modules (10\% photo-cathode coverage). Each segment is a 60m cube.}
\end{figure}

Such  a
detector could be operational within 10 years, with no major innovation or
development required. 
The baseline conceptual  design
of the UNO detector  is  shown  in  Figure~\ref{fig:UNODetector}.  The
detector  has  a  total 
(fiducial) mass of 648 (445) kton. As in \SK, 
the outer detector  region  serves  as  a
veto shield of 2.5 m depth, and is instrumented with  14,901  outward-facing
8" PMTs at a density of 0.33 PMT/m$^{2}$. The inner detector regions  are
viewed by 56,650  20"  PMTs. The  PMT  density  in  the
central sub-detector module is  chosen  to  give  40\%  photo-cathode  coverage
(equivalent to SuperK) and in the two optically isolated 
outer modules  to  be  10\%  each, allowing
sensitivity to a broad range of nucleon decay  and  neutrino  physics  while
keeping PMT costs under control.   

In
this configuration, the trigger threshold for the two wings would be  around
10 MeV deposited energy,  while the central module threshold is 5 MeV,
allowing efficient detection of 6 MeV $\gamma$s from
$p\rightarrow K^{+}\bar{\nu}$ decay,  precision  solar
neutrino studies and extraction of additional information on  core  collapse
from supernovae neutrinos, along with measurement of the $\nu_{\mu}$
and $\nu_{\tau}$  fluxes using neutral current excitation of oxygen.

With an outer veto detector and waveform-capture electronics, known
cosmic ray backgrounds even at modest depth (\about{}2,000\unit{\mwe{}})
will not compromise nucleon decay studies. However, less well understood
backgrounds such as cosmogenic fast neutrons could be a problem at
shallow depths.  The demands of a supernova relic
neutrino search and a solar neutrino physics program are more stringent, 
and would require a depth
of at least 3,000\unit{\mwe{}} to avoid unacceptable inefficiency or
background from muon-induced spallation products. This represents the minimum
acceptable depth for the detector, met by a number of potential sites. 
The Henderson Mine, for example, would allow
relatively low cost excavation at depths just below current commercial mining
operations, at about 4,000\unit{\mwe{}}.

\section{UNO Physics Capabilities}
\subsection{Nucleon Decay}

\begin{figure}
\vspace*{13pt}
\leftline{\hfill\vbox{\hrule width 5cm height0.001pt}\hfill}
         \mbox{\epsfig{figure=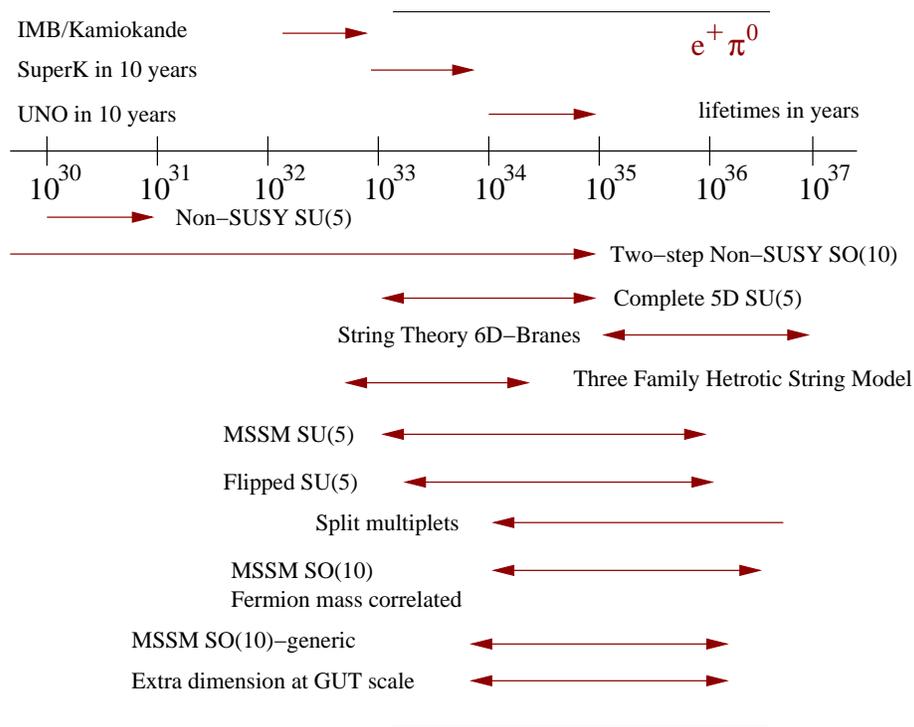,width=0.8\textwidth}}
\vspace*{30pt}
\leftline{\hfill\vbox{\hrule width 5cm height0.001pt}\hfill}
         \mbox{\epsfig{figure=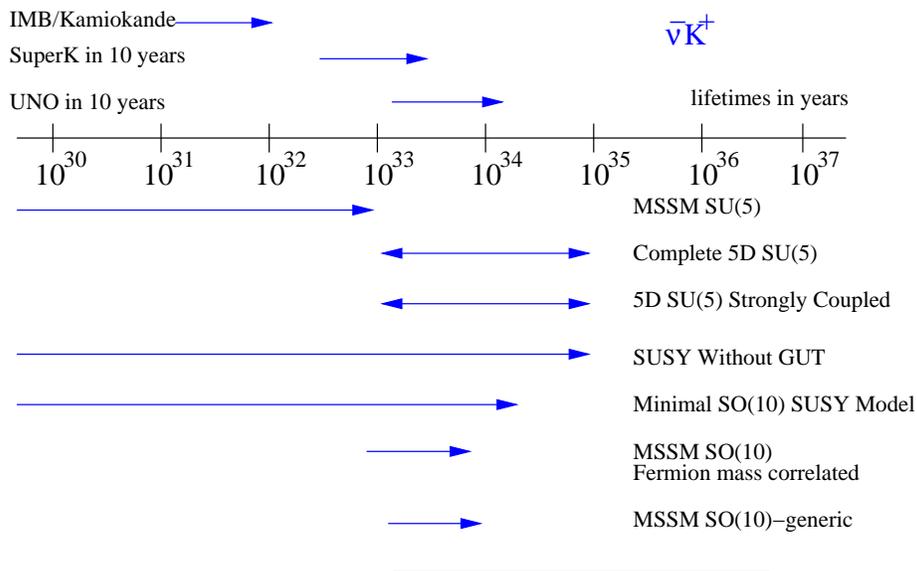,width=0.8\textwidth}}
\vspace*{0.5truein}		%ORIGINAL SIZE=1.6TRUEIN x 100% - 0.2TRUEIN
\leftline{\hfill\vbox{\hrule width 5cm height0.001pt}\hfill}
    \caption{\label{fig:PDecaySummary} Theoretical predictions of proton
      decay compared to experimental reach. Upper: $p\rightarrow e\pi^{0}$,
      Lower: $p\rightarrow\overline{\nu}K^{+}$.}
\end{figure}

Proton decay is a crucial prediction
of Grand Unification Theories of fundamental particles and forces.  Thus
the observation of proton decay would have far-reaching impact on our
understanding of nature at the highest energy scale. A wide range of
theoretical models predict nucleon decay (see
Figure~\ref{fig:PDecaySummary}).  

Background for nucleon decay is due primarily to atmospheric neutrino
interactions.  A small fraction of atmospheric neutrinos
lie in the proton decay candidate area on a plot of net momentum versus
visible energy. 

\begin{figure}
\vspace*{6pt}
\leftline{\hfill\vbox{\hrule width 5cm height0.001pt}\hfill}
         \mbox{\epsfig{figure=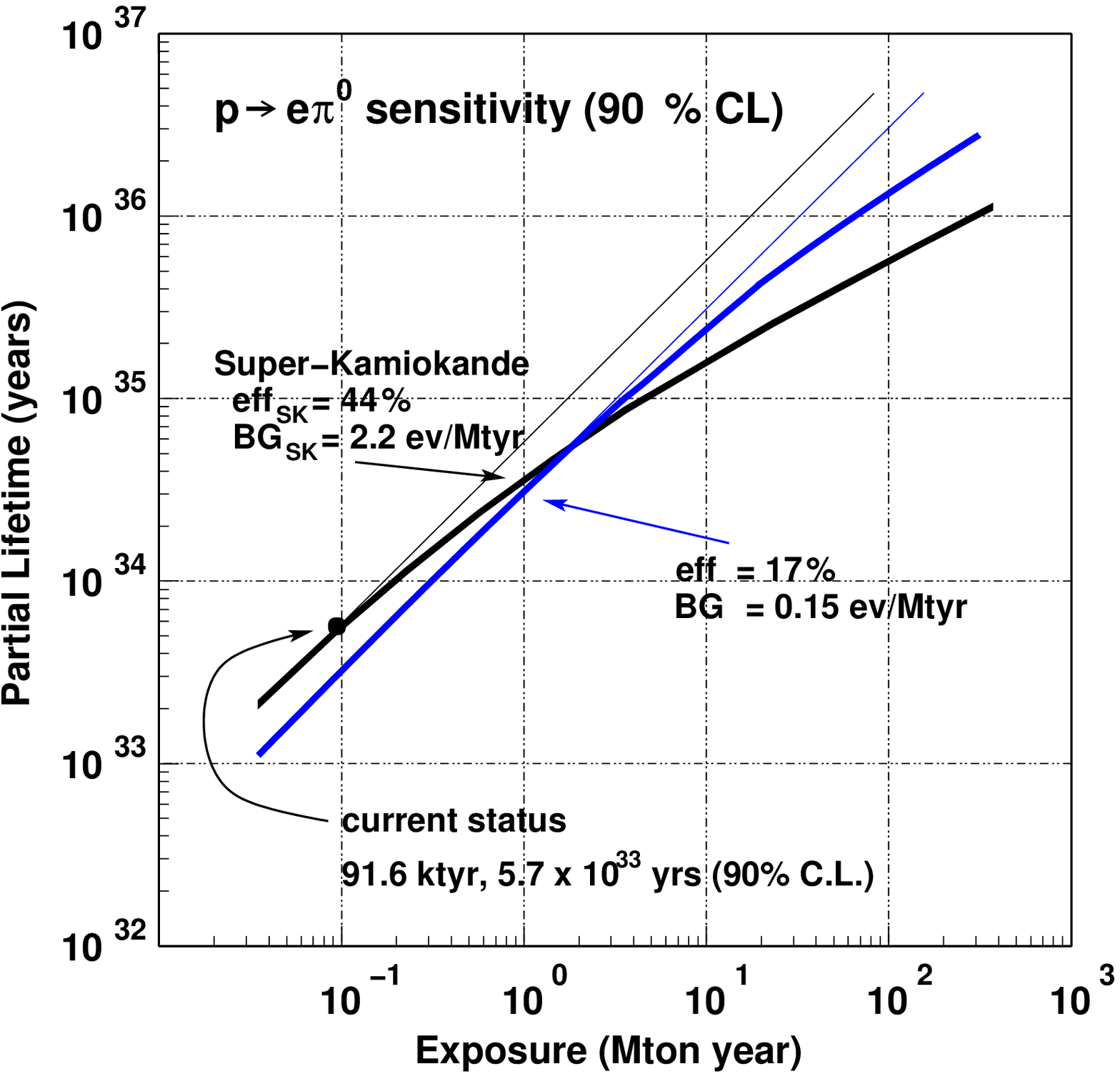,width=0.65\textwidth}}
\vspace*{6pt}
\leftline{\hfill\vbox{\hrule width 5cm height0.001pt}\hfill}
         \mbox{\epsfig{figure=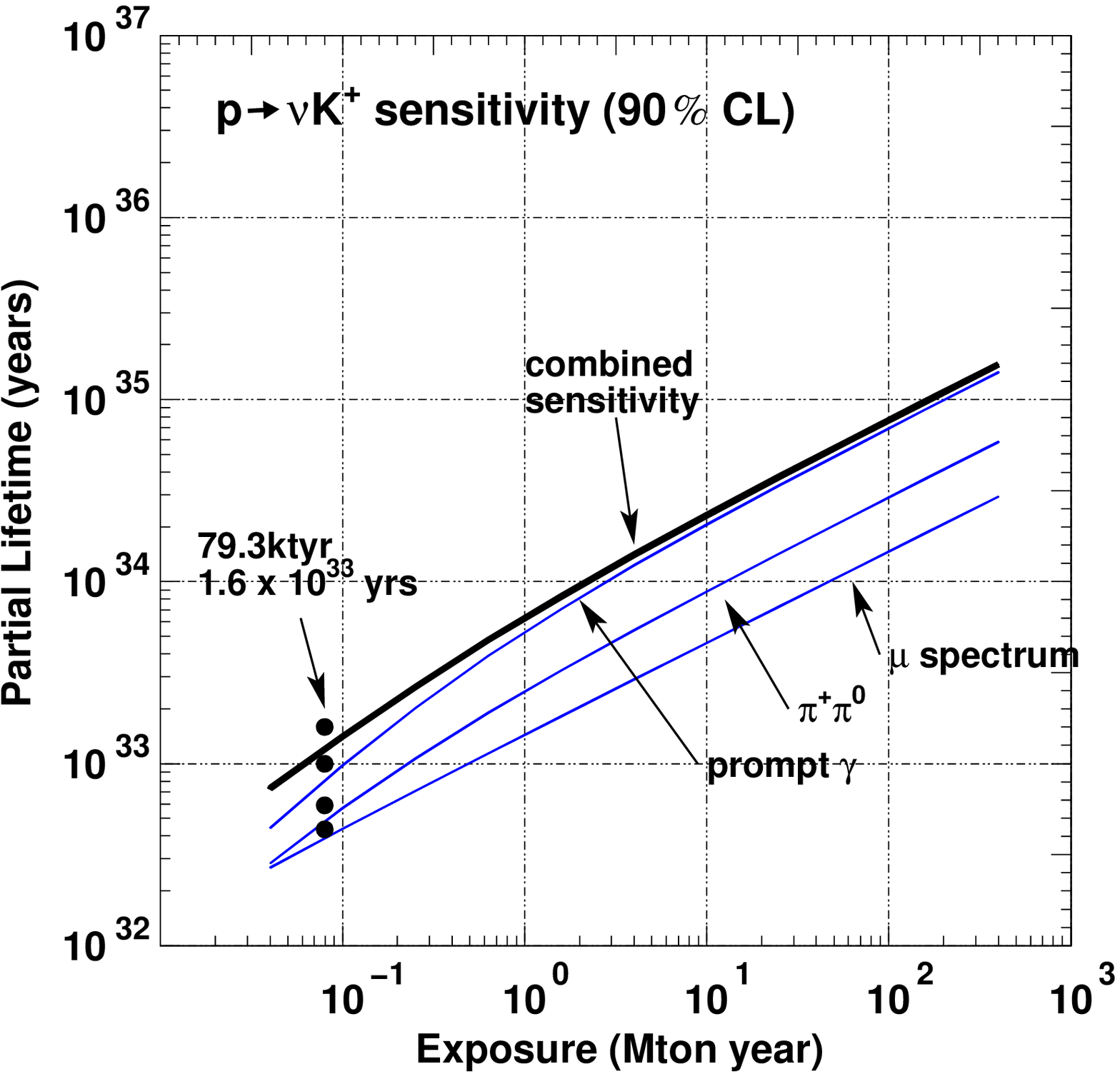,width=0.65\textwidth}}
\vspace*{6pt}		%ORIGINAL SIZE=1.6TRUEIN x 100% - 0.2TRUEIN
\leftline{\hfill\vbox{\hrule width 5cm height0.001pt}\hfill}
  \caption{\label{fig:PDecayExpected}
UNO sensitivity to the partial lifetimes of \peppo{} (upper) 
and \pnukp (lower)
as a function of total exposure at 90\% confidence level.}
\end{figure}

\begin{figure}
\vspace*{13pt}
\leftline{\hfill\vbox{\hrule width 5cm height0.001pt}\hfill}
         \mbox{\epsfig{figure=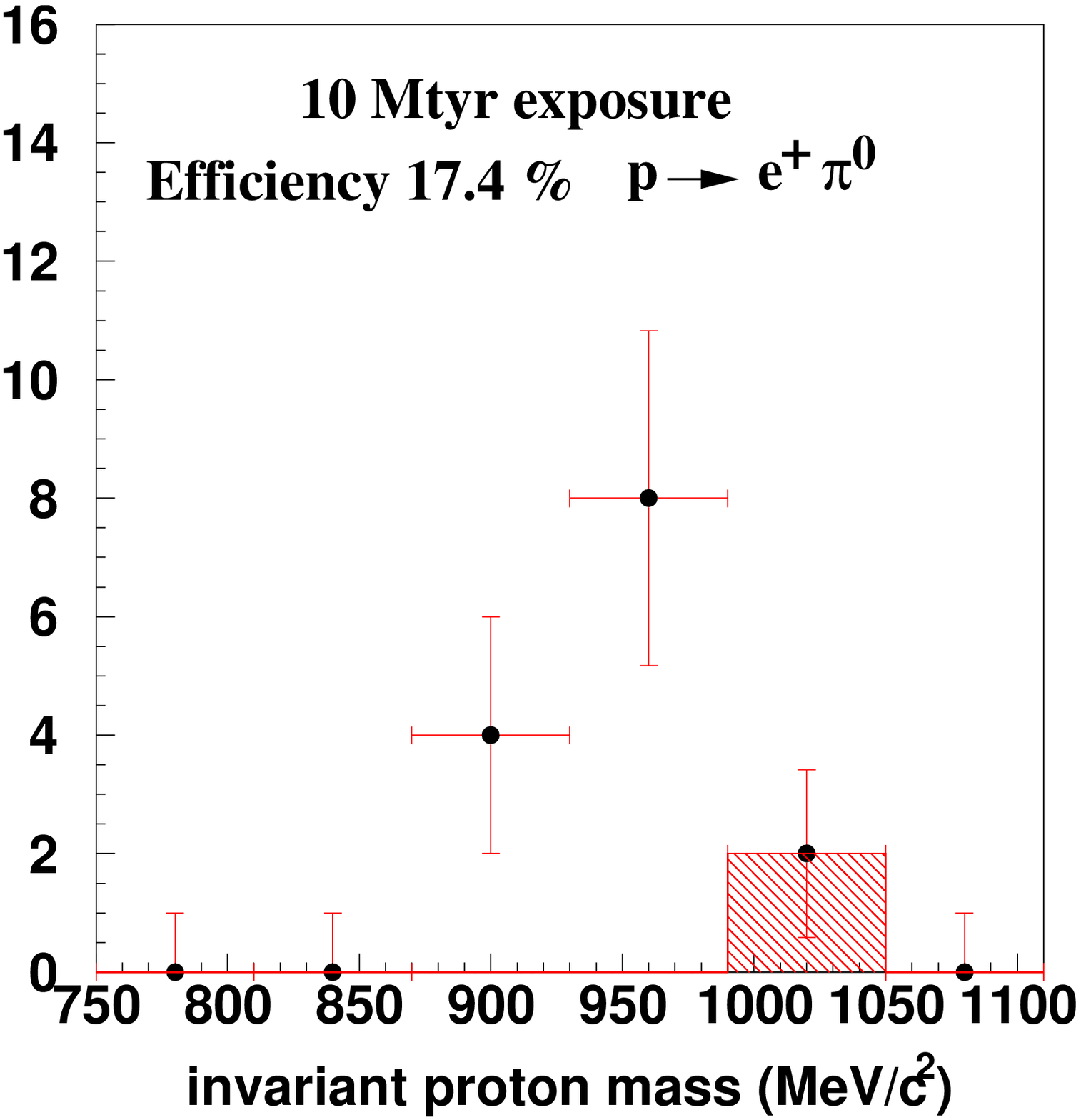,width=0.8\textwidth}}
\vspace*{1.4truein}		%ORIGINAL SIZE=1.6TRUEIN x 100% - 0.2TRUEIN
\leftline{\hfill\vbox{\hrule width 5cm height0.001pt}\hfill}
  \caption{\label{fig:PDecayMass} Expected invariant mass distribution of
   \peppo{} candidate events passing all selection criteria.  
Detector exposure of
    5\unit{\megatonyr{}} and partial proton lifetime of 5\E{34}
    years are assumed. The hatched histograms represent the backgrounds.}
\end{figure}

To estimate UNO sensitivity for nucleon decay, a
20\unit{\megatonyr{}} equivalent sample of atmospheric neutrino background
events and large samples of nucleon decay candidate events have
been simulated and reconstructed using SuperK
neutrino interaction and detector simulation tools with UNO photocathode
coverage (40\% and 10\%).  The results
are shown in
Figure~\ref{fig:PDecayExpected}. With no signal events, five
years of UNO operation will extend the lifetime limit for two
``benchmark" decay modes (\peppo{} and \pnukp) by roughly an order
of magnitude over present limits, to \about{}5\E{34}\unit{\yr{}}
and \about{}10$^{34}$\unit{\yr{}}, respectively, and the 10$^{35}$ yr 
level is reached after 13 years (at 6\unit{\megatonyr{}}).  
Figure~\ref{fig:PDecayMass}
shows expected invariant mass distribution for \peppo{} candidates
with 40\% PMT coverage and a 5\unit{\megatonyr{}} exposure,
assuming partial proton lifetimes of 5\E{34}\unit{\yr{}}.

\subsection{Supernova Neutrinos}

The number of neutrino events recorded by UNO from a supernova
collapse will be so great that detailed analyses of spectra and time
structure will be possible. For a Type II supernova near the center
of our Galaxy, at 10 kpc distance, a total
of \about{}140,000 neutrino events are expected.

\begin{figure}
\vspace*{13pt}
\leftline{\hfill\vbox{\hrule width 5cm height0.001pt}\hfill}
         \mbox{\epsfig{figure=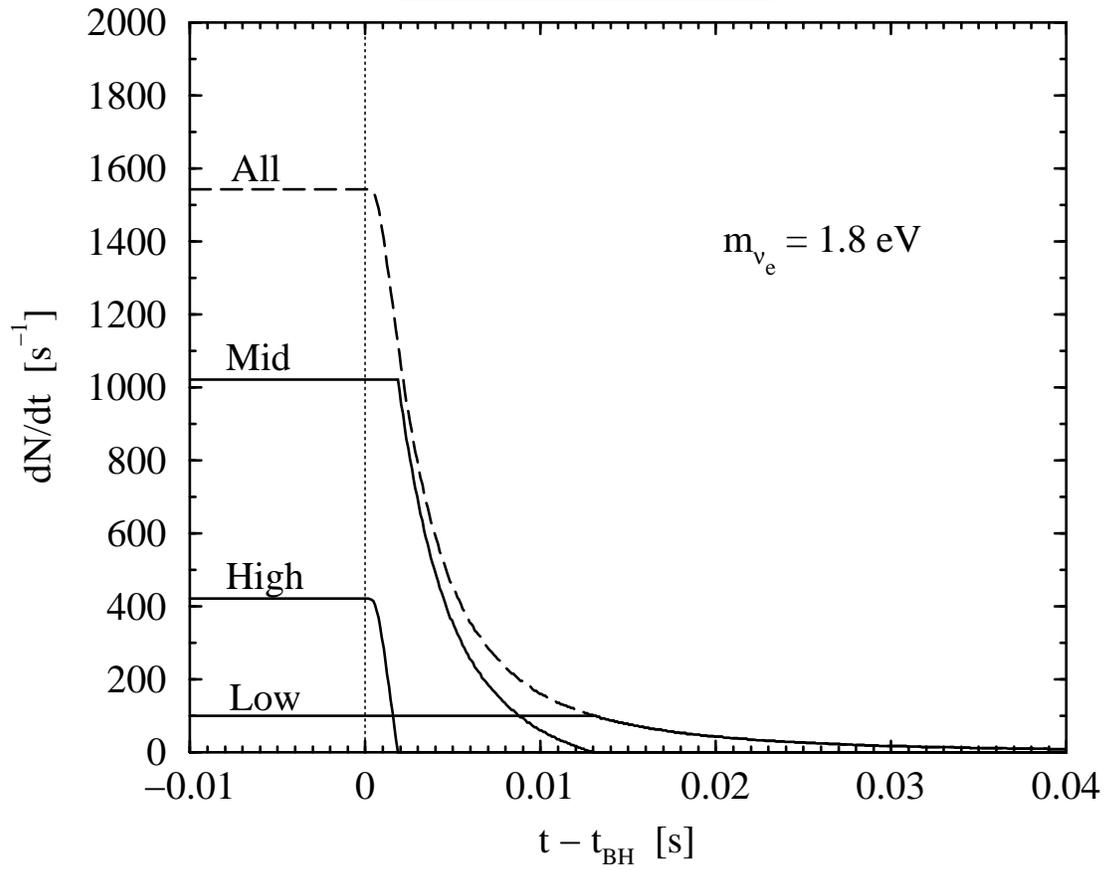,width=0.95\textwidth}}
\vspace*{0.5truein}		%ORIGINAL SIZE=1.6TRUEIN x 100% - 0.2TRUEIN
\leftline{\hfill\vbox{\hrule width 5cm height0.001pt}\hfill}
    \caption{\label{fig:BlackHole} Detection of Black Hole Formation
      and $\nu_e$ Mass Determination.  The neutrino data is 
      subdivided according to energy
      range to provide a sharp ``time zero" for the black hole formation
      (high energy) and well-defined delayed arrival times (middle energy).
      This delay is directly related to the mass of $\nu_e$.  (The event
      rates shown correspond to detection in SuperK of a supernova
      at 10 kpc from Earth.)}
\end{figure}

Such a high-statistics observation will 
allow investigation of the millisecond scale behavior
of the light curve, especially at early times, providing information on
core collapse mechanisms. In addition, the event rate will be high enough to
study the late time
behavior of the light curve. The event rate from
a supernova should gradually decrease over tens of seconds, but if a black
hole forms during a supernova explosion (based on recent calculations, with
probability
about 50\%), the neutrino flux will be sharply cut off as the event horizon
envelops the neutrino-sphere of the imploding star\cite{ref:BlackHole} (see
Figure~\ref{fig:BlackHole}). Additional results which 
such a large statistics database makes possible include:

\begin{itemize}
  
\item A calorimetric measurement of the total energy radiated in neutrinos
  will yield the neutron star binding energy\cite{ref:Barger}.
  To a good approximation for
  most equations of state, the dimensionless binding energy is given by
  $BE/M \sim \frac{3}{5}(\frac{GM}{Rc^{2}})(1-\frac{GM}{2Rc^{2}})^{-1}$
  constraining the mass and radius of the remnant neutron star.

\item Simultaneous flux and spectral information at each epoch, combined
  with simulations, can yield the angular size
  $\frac{R_{\infty}}{D}=\frac{R(t)}{D\sqrt{1-(2GM/R(t)c^{2})}}$ of the
  proto-neutron star at each epoch. If the distance D to the supernova can
  be otherwise measured, this would result in an independent measurement of
  mass and radius. Combined with measurement of the total radiated neutrino
  energy, which refers to the late-time radius
  $R_{\infty}=R(t\rightarrow\infty)$, both the mass and radius can be
  inferred.
  
\item Should the flux suddenly disappear before passing below the threshold
  of detection, one could infer that the proto-neutron star was metastable
  and collapsed into a black hole as mentioned above\cite{ref:BlackHole}.
  Deleptonization of the star could result
  in a new phase appearing, such as hyperons, a kaon or pion condensate, or
  quark matter, that effectively reduces the maximum mass below the star's
  actual mass.
  
\item Details of the neutrino flux curve and time evolution of the average
  neutrino energy (i.e., when the average energy peaks, when neutrino
  transparency sets in, etc.) will additionally constrain opacities and the
  proto-neutron star mass\cite{ref:Takahashi}.
  
\item Relative proportions of \nue{}, \nuebar{}, etc., will further test
  simulations and ought to reveal details of neutrino oscillations.

\item If the supernova's sky location at burst is such that neutrinos pass
through a substantial thickness of the 
  Earth, matter effects may allow us to improve our knowledge of neutrino
  oscillation parameters\cite{ref:Takahashi1}.
 
\end{itemize}

UNO is sensitive to supernovae out to about 1 Mpc, encompassing our local group of
galaxies, notably M31 (the Andromeda Galaxy). For a supernova in Andromeda, the 
total number of events would only be on the order of 10, but
having this additional reach will allow UNO to observe supernovae three
times more frequently than detectors limited to our own galactic
neighborhood.  Moreover, since telescopes on Earth view M31
face-on, the chance of observing
the optical counterpart for a neutrino burst is about three times greater
than in the obliquely-viewed, dust-obscured plane of the Milky Way. 

\subsection{Supernova Relic Neutrinos (SRN)}

\newcommand{\nuflux} {\ensuremath{\nuebar{}\!\cdot{}\!
    \cm{}^{-2}\!\cdot{}\!\mbox{s}^{-1}}}

Supernova relic neutrinos are a low intensity isotropic background of
cosmic neutrinos originating from core-collapse supernovae, summed over the 
history of the Universe.
Recently, \SK ~published a search for SRN using
\nuebar{}p\goesto{}n$e^{+}$ interaction in the energy range E $>$
19\unit{\MeV{}}.  At these energies, the predicted fluxes are
0.20-3.1\unit{\nuflux{}}\cite{ref:SKSNR}. In the absence of a signal,
a 90\% C.L.
limit of 1.2\unit{\nuflux{}} was set.  While this limit 
eliminates some theoretical models, a factor of six reduction is
needed to challenge all current predictions.
If UNO were built at a depth of 4,000\unit{\mwe{}}, 
it would reach this level within six
years.

\subsection{High Energy Neutrino Astrophysics}

No direct observation of a non-transient neutrino source
other than the Sun has yet been reported, despite the fact that neutrinos
must be produced by the same meson decay processes that produce high energy
gamma rays, and thus in proportionate abundance. However, UNO
will provide enormous effective area for detecting upward-going
muons, which represent the highest-energy sample of neutrino
interactions the experiment can collect. As in \SK ~and MACRO, 
searches can be conducted for point
sources of high energy neutrinos such as AGNs, neutrinos from GRBs and WIMP
annihilations at the center of the Earth, the Sun and our Galaxy.
UNO can fully contain muons with energies up to $\sim$40 GeV, and can
observe through-going muons with energies of hundreds of GeV. Thus, it can
provide means to search for astrophysical neutrino sources in the range 
between that covered by existing underground detectors, 
and large planned under-ice and 
underwater ultra-high energy neutrino detectors.

\subsection{Long Baseline Neutrino Oscillation Experiments}

UNO is well-suited as the far detector for future long-baseline neutrino
oscillation experiments.  The neutrino source could be either a
high-intensity conventional beam (a ``super-beam"), or a pure
\nue{} (\nuebar{}) beam from the beta decay of short-lived isotopes
using a relatively low energy storage ring (a ``beta-beam").

There have been a number of proposals for a long baseline neutrino
oscillation using a super-beam.  Published case studies, including one on a
130\unit{\kilometer} baseline experiment using the CERN SPL and UNO at
\Frejus{}\cite{ref:UNO}\cite{ref:Gomez}, and another on the 
T2K Phase II, using a 4 MW proton driver and
Hyper-Kamiokande with a 295\unit{\kilometer{}} baseline, showed that
CP violation in the lepton sector could be observed\cite{ref:JHFHK}.

A recent study performed at Brookhaven National Laboratory 
proposes a neutrino super-beam using
an upgraded AGS optimized for a very long baseline experiment
\cite{ref:BNL}. With the proposed beam aimed
at a 500\unit{\kiloton{}} water Cherenkov detector at distances over
2,500\unit{\kilometer{}}, sensitivities to oscillation parameters would be: 
(1) Measurement of $\sin^{2}\theta_{13}$ to below 0.005; (2)
Determination of the sign of $\Delta m^{2}_{31}$; (3) Measurement of
$\sin\delta$ (and $\cos\delta$) to about 25\% level thus determining
J$_{CP}$ and $\delta$; (4) Measurement of $\Delta m^{2}_{21}$ and
$\theta_{12}$ from the $\numu{}\leftrightarrow\nue{}$ oscillation in an
appearance mode. UNO built at Henderson (2,760\unit{\kilometer{}}),
or Homestake (2,540\unit{\kilometer{}})
would be an ideal far detector for such a beam.

Another study performed at Fermilab focuses on an off
axis detector along the NuMI beamline\cite{ref:NuMI}. Preliminary results
show that an UNO-scale water Cherenkov detector located
15\unit{\millirad} off axis at
735\unit{\kilometer{}} would have the sensitivity to exclude
$\sin^{2}\theta_{13}$ down to about 0.006 after 5 years.  
With an accelerator upgrade to
neutrino beam intensity together and build an anti-neutrino beam, it would be
possible to search for CP violation as well as extending measurements on $\theta_{23},
\Delta m^{2}_{23}, \theta_{13}$ and $\Delta m^{2}_{13}$\cite{ref:fermi}.

\subsection{Atmospheric Neutrinos}

\SK ~has presented compelling evidence for
atmospheric muon neutrino disappearance\cite{ref:SKatm}, and recently
showed,
using data samples selected for good resolution in L/E, 
that a dip corresponding to an oscillation minimum is being
observed\cite{ref:SKLoE}.

\begin{figure}
\vspace*{13pt}
\leftline{\hfill\vbox{\hrule width 5cm height0.001pt}\hfill}
         \mbox{\epsfig{figure=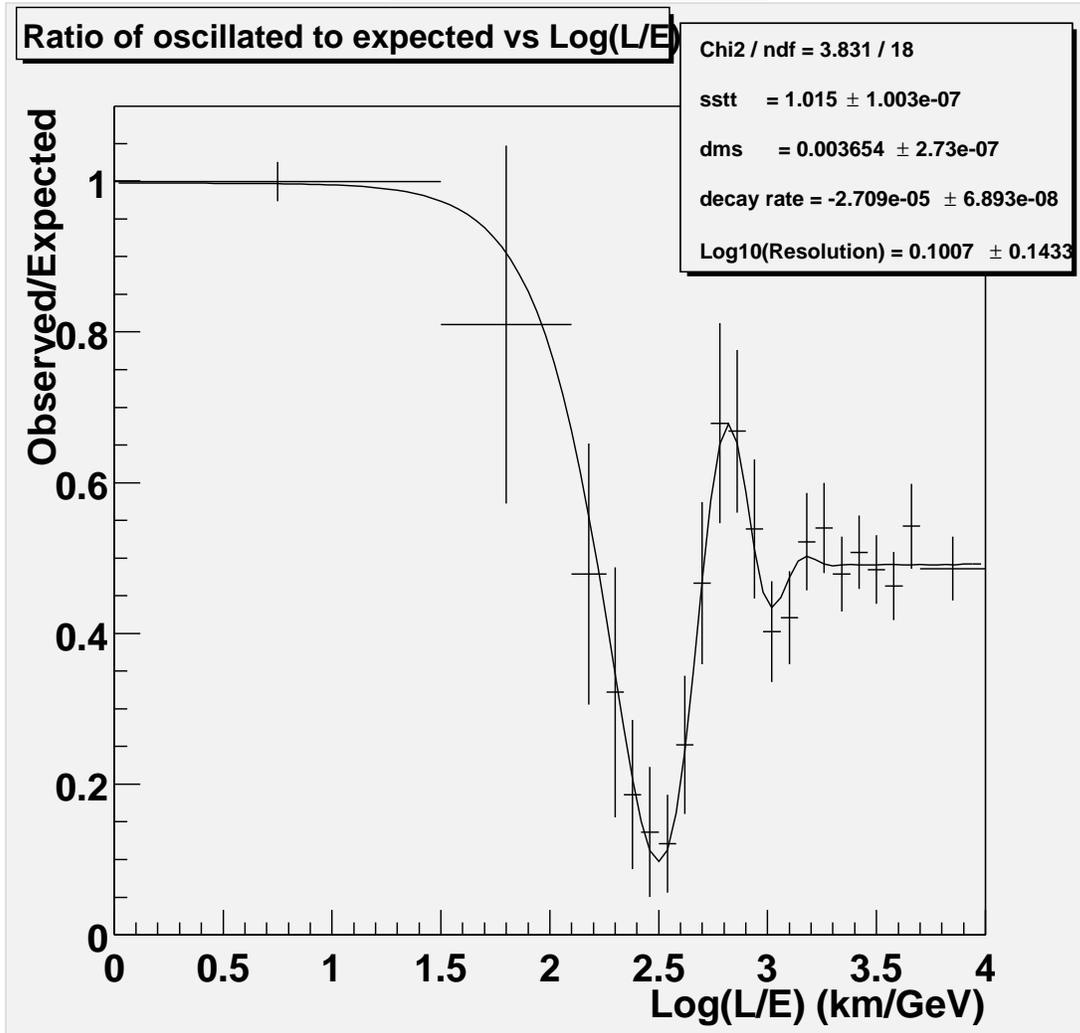,width=0.95\textwidth}}
\vspace*{1.4truein}		%ORIGINAL SIZE=1.6TRUEIN x 100% - 0.2TRUEIN
\leftline{\hfill\vbox{\hrule width 5cm height0.001pt}\hfill}
  \caption{\label{fig:LoE} The ratio of the oscillated muon  event  rate
    to the expected rate as a function of L/E assuming a
    2830\unit{\kilotonyr{}} exposure (a \about{}7\unit{\yr{}} UNO run).
    The oscillated flux assumes the parameters are $\Delta
    m^2=0.003 \unit{\eVs{}}$, and sin$^{2}2\theta$=1.}
\end{figure}

However, the possibility 
that the observed behavior is due to some phenomenon other than
neutrino oscillation has not yet been excluded.
In fact, several models have been proposed where the
expected disappearance of \numu{} is of the form $e^{-\alpha L/E}$ with
$\alpha$ determined by the model. The multiple oscillation 
sinusoidal pattern expected from neutrino oscillation can be
established conclusively by high-statistics 
measurements of atmospheric neutrinos in a
larger detector. In \SK, 
the detector dimensions are too small to efficiently contain muons with
energies above several GeV, which is crucial for observing oscillatory
behavior in atmospheric neutrinos.  UNO can contain muons with
energies up to ~40\unit{\GeV{}}. Figure~\ref{fig:LoE} shows the sensitivity
expected on the ratio of signal to expectation where the oscillation
parameters have been assumed to be $\Delta m^{2}=0.003\unit{\eVs{}}$ and
sin$^{2}2\theta$=1.  

New physics can be extracted from high statistics atmospheric neutrino data
by invoking ``global" fits for three-generation neutrino mixing.  For
example, global fits will establish (or at least indicate) new,
constraining limits for possible sub-dominant contributions from sterile
neutrinos. In addition, UNO can search for amplification of sub-dominant
\numu{} to \nue{} oscillation resulting from matter resonances in the
Earth\cite{ref:Shiozawa}. 

SuperK data showed $\numu{}\leftrightarrow\nutau{}$ provides an
explanation for the atmospheric neutrino zenith angle distributions.
Assuming two component, full mixing, with \SK  ~best-fit parameters, 
approximately one \nutau{}
charged current (CC) event is expected per kiloton-year of exposure. Thus we
would expect about 400 \nutau{} CC events per year in UNO, providing
more than a three standard deviation excess after one year exposure.

\subsection{Solar Neutrinos}

Only a detector the size of UNO will have an event
rate that is sufficiently high to detect with statistical confidence the
day-night effect in solar neutrinos, the characteristic
signal of matter-induced neutrino oscillations (MSW effect). The UNO 
central module, with 40\% photo-cathode coverage, will have a 
threshold of 6\unit{\MeV{}} for solar neutrinos. The best-fit LMA
solution predicts a 2\% day-night difference in event rates, which can be
observed as a 4$\sigma$ effect with UNO in approximately ten years.
The total event rate for $^8$B solar neutrinos in UNO will be
about 3\E{4} events per year.

The hep process, $^3He + p \to ^4He + e^+ + 
\nu_e $,  is a rare branch of the $pp$ chain in the Sun, yielding the 
highest energy solar neutrinos, up to 19 MeV, but under the Standard Solar Model 
the hep flux should be three orders of magnitude smaller than that 
of $^8$B neutrinos. Super-Kamiokande performed a study of hep neutrinos 
\cite{ref:Smy02} 
and found a barely significant signal 
in 5 years of data. UNO's large statistics will provide
an order of magnitude improvement in the search for hep neutrinos.

\section{Strawman site for UNO: Henderson Mine}

\begin{figure}
\vspace*{13pt}
\leftline{\hfill\vbox{\hrule width 5cm height0.001pt}\hfill}
         \mbox{\epsfig{figure=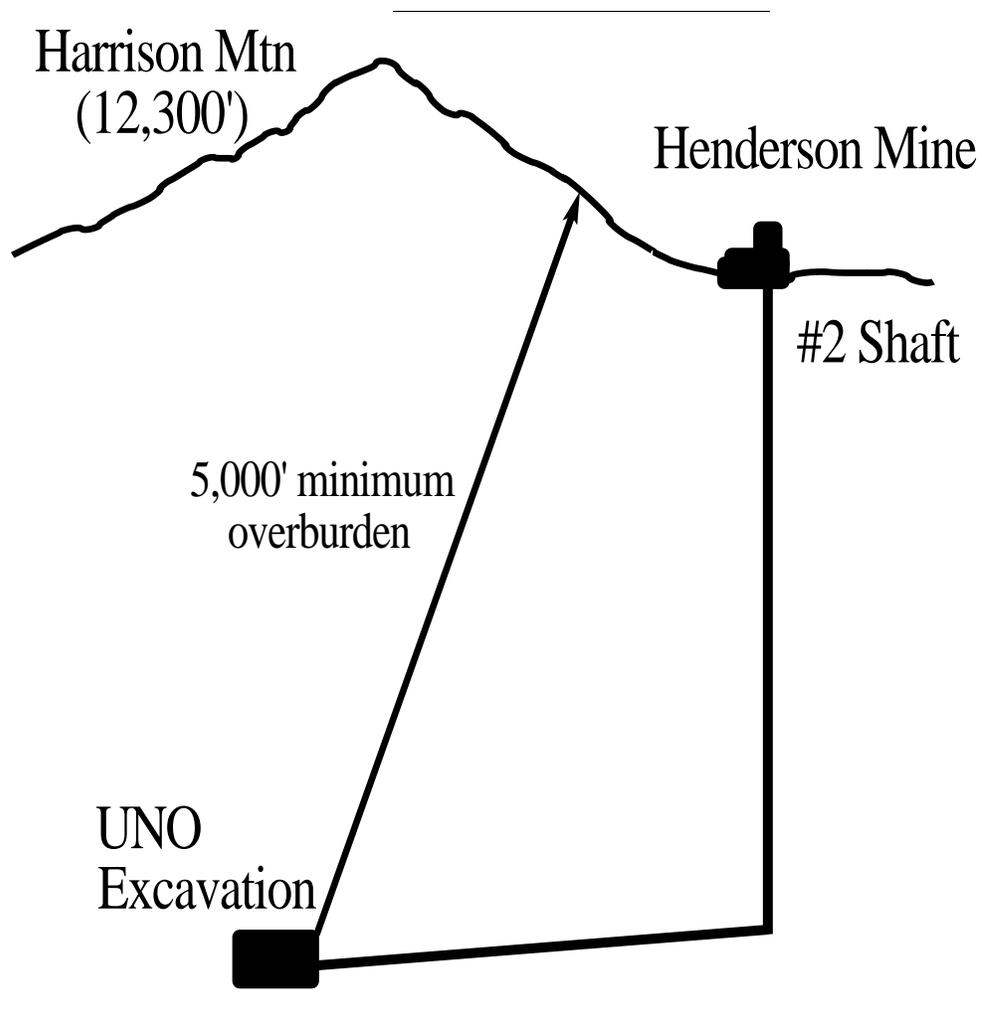,width=0.9\textwidth}}
\vspace*{1.4truein}		%ORIGINAL SIZE=1.6TRUEIN x 100% - 0.2TRUEIN
\leftline{\hfill\vbox{\hrule width 5cm height0.001pt}\hfill}
  \caption{\label{fig:vertical-section} Vertical section through Harrison
    Mountain showing a possible location for the UNO cavity, 
easily reached from operating areas in the Henderson mine, with overburden
approximately 4000\unit{\mwe{}}.}
\end{figure}

The Henderson mine, located near Empire, Colorado, in 
the Rocky Mountains, is operated by
Climax Molybdenum Company, a subsidiary of Phelps Dodge
Corporation. The Henderson
mine produces molybdenum ore by an underground mining
method known as panel caving. The mine is located 
80.5 km (50 mi) west of Denver, Colorado, and can be
reached in about 1.5~hr driving time from Denver International
Airport.
The surface facilities lie 3170~m (10,400~ft)
above sea level, just east of
the North American 
Continental Divide, which separates the Mississippi River watershed
from the Pacific watershed. 
The approximately 11.7 km$^2$ (2900 acre) parcel of land
containing the Henderson orebody, located underneath Red Mountain, as well
as the proposed UNO site, located underneath Harrison Mountain
(Fig~\ref{fig:vertical-section}), is entirely privately owned by Climax
Molybdenum Company.  Additionally, the 52- km$^2$ (12,800 acre) mill site
located near Kremmling, Colorado is also entirely owned by the company.

The mine infrastructure is engineered to support
production in excess of
about 10 million tons per year, easily making the Henderson Mine
one of the 10 largest underground mining operations in the world today.
The mine is currently producing about 21,000 tons of raw ore per
day. Material excavated from the UNO cavity would
be a negligible percentage of the debris produced by normal mine
operations, for which disposal is already planned and permitted.
It is estimated that the mine has ore reserves for about
twenty more years of production. 
The rock is very ``competent"
(high strength) granite with compressive strengths ranging from
100 to 275 Mpa (14,500 to 40,000 psi). 

At Henderson Mine, UNO facilities would be located under Harrison
Mountain.  Previous geologic studies have
indicated this area to be free of economically interesting mineral
deposits. Coring conducted in 2004 indicates the UNO cavity would be in
competent Precambrian Silver Plume Granite. 
The mine is accessed from the surface by an 8.53~m (28 foot)
diameter personnel and material shaft that extends down to the 7500
level. (Note: In US mining parlance, levels are referred to by their
altitude in feet above sea level, not by their depth below the surface.)
 The shaft cages can transport up to 200 people at a time,
and the trip from the surface to the 7500 level takes about 5
minutes. The cage can accommodate loads with maximum dimensions
2.6~m wide, 7.1~m long, and 3.9~m high, weighing up to 30 tons.  
Loads of up to 50 tons can be carried with special preparation.

At the 7500 level, ore is loaded onto an
underground conveyor, and transported approximately horizontally
16.1 km (10 mi) under the continental divide 
and out of the mountain by the longest single flight 
conveyor system in
the world.  It is then transferred to another
6.4 km (4 mi) long overland conveyor that transports the
ore to stockpiles for subsequent processing. Tailings are placed in large
containment areas that will be reclaimed and re-vegetated when the
mine is closed.  The operating permit allows for the deposition of
in excess of 340 million tons of mill tailings. The UNO Cavity tailings
will be a negligible addition to these planned deposits.
Henderson is the second largest consumer of electricity
in  Colorado, with a permanent substation located on the
property. Henderson
Mine was constructed in the early 1970s, and facilities were
significantly upgraded in 2001, so Henderson is an exceptionally modern, 
safe and well equipped 
site for UNO (for example, fiber-optic data lines are in place throughout the 
underground operations area). The mine meets or exceeds all current US safety standards.

The proposed location of the UNO cavity is directly below
the  summit of Harrison Mountain, elevation 3750~m (12,300~ft).
Because of the availability of the large capacity shaft and mine
tunnels developed as part of the infrastructure for the Henderson
mining operation, the amount of new access
tunneling required will be minimal.

A vertical section through Harrison Mountain
showing the approximate location of the UNO cavity relative to the
surface is given in Figure~\ref{fig:vertical-section}. 
The proposed UNO room will have dimensions of 60~m wide, 60~m
high, and 180~m long (200 $\times$ 200 $\times$ 600~ft). Including
the arched opening over the room, it is estimated that about 1
million m$^3$ (35,315,000~ft$^3$) of rock will have to be removed,
or about 2,780,000 tons. Rock will be excavated and removed via the
Henderson ore transportation system at an estimated cost of
about US\$7.50/ton (in 2004), an extremely low rate. 

With a minimum 1500m (5,000' or approximately 4000\unit{\mwe{}}) 
overburden at the proposed location of the
UNO excavation
cavity, the Henderson Mine is deep enough to meet the low cosmic
ray background levels required for UNO and most other proposed
underground experiments. Greater depths are readily achievable at
relatively low cost.

It is remarkably lucky that such a highly suitable site is
in operation 
only a few kilometers away from a major highway and about 1 hr from a
large metropolitan area, which provides
technical industries, research
universities, and a major international airport. Since 
the mine is a significant local employer and
has been in operation for over 30 years, there are no
unresolved community-relations or environmental issues.
State and local government officials
and community representatives are enthusiastic about the possibility of
a long term underground laboratory at Henderson, and lent support to a site 
proposal submitted
for the US Deep Underground Science and Engineering Laboratory (DUSEL)
\cite{ref:HUSEP05}. It was recently announced that Henderson
has been selected by NSF as one of 
two finalists (with the Homestake mine) for the DUSEL site. During
the coming months, a fully detailed construction proposal for
the underground laboratory will be prepared.

\section{Conclusion}
In late Summer 2005, the UNO Collaboration will request support from
US funding sources for 
R\&D needed to prepare a full construction proposal about 2 years later.
The research effort will include studies of mine engineering and
cavity construction methods, detector development, and physics simulations. 

Construction of UNO will require about 10 years, including two
years of contingency in the excavation schedule and one year of
contingency in the overall schedule. PMT delivery is a limiting factor, but
the rate could be increased with additional cost.

UNO brings together a remarkable collaboration of physicists and engineers from
a wide range of fields\cite{ref:UNOweb}. There is increasing consensus in
the particle physics community that a megaton scale water Cherenkov detector
will be an essential facility in coming years. At future Neutrino Telescopes workshops,
we hope to report on significant additional progress.

\section{Acknowledgements}
Thanks are due to Chang-Kee Jung for reviewing this paper, and to many UNO colleagues
for preparing the illustrations and information used. All errors are the responsibility of 
the author alone. Finally, special thanks are due to Milla Baldo-Ceolin for her
extraordinary patience!

%add bibliography here

\end{document}